\documentclass[twoside,10pt,a4paper]{newFNLstyle_mod}

\usepackage{epsfig} 
\usepackage{epsf} 
\usepackage[english]{babel} 
\usepackage{latexsym} 
\usepackage{amsmath} 
\usepackage{amssymb} 
\usepackage{graphicx} 
\usepackage{subfigure} 

\newcommand{\be}{\begin{equation}} 
\newcommand{\ee}{\end{equation}} 
\newcommand{\bea}{\begin{eqnarray}} 
\newcommand{\eea}{\end{eqnarray}} 
\newcommand{\h}{\mathcal{H}}
\newcommand{\ave}[1]{\langle #1 \rangle}

\begin{document} 

\volnumpagesyear{0}{0}{000--000}{2007}
\dates{-}{-}{-}

\title{Limited resolution and multiresolution methods in complex network community detection} 
\authorsone{Jussi M. Kumpula$^{1,*}$, Jari Saram\"{a}ki$^1$, Kimmo Kaski$^1$,  and J\'{a}nos Kert\'{e}sz$^{1,2}$ }
\affiliationone{
$^1$Laboratory of Computational Engineering, Helsinki University of Technology, P.O. Box 9203, FIN-02015 HUT, Finland; \\ 
$^2$Department of Theoretical Physics, Budapest University of Technology and Economics, Budapest, Hungary 
}
\mailingone{*e-mail: jkumpula@lce.hut.fi}

 
\maketitle 

 \pagestyle{myheadings}

\begin{abstract}
Detecting community structure in real-world networks is a challenging problem. Recently, 
it has been shown that the resolution of methods based on optimizing a modularity 
measure or a corresponding energy function is limited; communities with sizes below some 
threshold remain unresolved. One possibility to go around this problem is to vary the 
threshold by using a tuning parameter, and investigate the community structure at 
variable resolutions. Here, we analyze the resolution limit and multiresolution 
behavior for two different methods: a $q$-state Potts method proposed by Reichardt
and Bornholdt, and a recent multiresolution method by Arenas, Fern\'andez, and G\'omez. 
These methods are studied analytically, and applied to three test networks using 
simulated annealing.  
\end{abstract}

\selectlanguage{english}

Networks consisting of nodes and links are an efficient way to represent and study a large variety of technological, biological and social complex systems~\cite{RefWorks:72,RefWorks:77}.  
Usually the functionality of these systems is of central interest, which, on turn, is closely related to the structure of the corresponding networks. In particular, substructures called \emph{modules} or \emph{communities} are abundant in networks.
These communities are, loosely speaking, groups of nodes that are densely interconnected but only 
sparsely connected with the rest of the network~\cite{RefWorks:76,RefWorks:75,RefWorks:67,RefWorks:52} 
-- consider, \emph{e.g.}, groups of individuals interacting with each other in social networks, 
or functional modules in metabolic networks.  As communities are supposed to play a special role in the often stochastic dynamics of the systems under consideration, their identification is crucial. Thus, reliable and computationally tractable methods for detecting them in empirical networks are required.
 
Several  methods and algorithms have been developed for community detection~\cite{RefWorks:53,RefWorks:70}. 
One popular class of methods is based on 
optimizing a global quality function called modularity~\cite{RefWorks:46}, 
or a closely related Hamiltonian~\cite{RefWorks:50},   
which contains the modularity as a special case. 
The related methods are
computationally demanding, especially for large networks, but various approximative algorithms
exist~\cite{RefWorks:40,RefWorks:49,RefWorks:46,RefWorks:68,RefWorks:65}. 
For many test networks, these methods have been shown to perform well~\cite{RefWorks:53,RefWorks:69}. 
However, it has recently been shown that the resolution of the modularity based methods is intrinsically limited, that is, modularity optimization fails to find 
small communities in large networks -- instead,  small groups of connected nodes turn out merged as 
larger communities~\cite{RefWorks:104}. For the Hamiltonian-based method, there is also a resolution limit due to similar 
underlying reasons~\cite{RefWorks:105} though this method contains a tuning parameter which can be 
used to study communities of different sizes. Recently, Arenas \emph{et al.} proposed a modification 
of the modularity optimization method which also provides a parameter that can be used to probe the 
community structure at different resolutions. 
Here, we compare these two methods and their resolutions analytically, pointing out similarities 
and differences. Subsequently we apply them to several test networks using optimization by simulated 
annealing.

We start by briefly reviewing the concept of modularity, introduced by Newman and Girvan~\cite{RefWorks:46}.
The modularity $Q$ is defined as follows 
\be 
Q=\frac 2 K \sum_{s=1}^m ( l_{ss} - [l_{ss}] ), 
\label{eq:newman3} 
\ee
where $K$ is the degree sum of the network,
$l_{ss}$ is the number of links in community $s$,
$ [l_{ss}] \equiv  K_s^2/2K$ is the {\it expected} number of links inside community $s$, given 
that the network is random, and $K_s$ 
is the sum of the degrees of nodes in community $s$.
In modularity optimization, the goal is to assign all nodes into communities  
such that $Q$ is maximized. 
 
The Hamiltonian-based method introduced by Reichardt and Bornholdt (RB) 
is based on considering the community indices of nodes as spins in a $q$-state 
Potts model, such that if the energy of such as system is minimized, groups of nodes with dense internal  
connections should end up having parallel spins~\cite{RefWorks:50}.  
The Hamiltonian for the system is defined as follows: 
\be 
\h = - \sum_{s=1}^m \left( l_{ss} -\gamma [l_{ss}]_{p_{ij}} \right), 
\label{eq:rb1} 
\ee 
where $[l_{ss}]_{p_{ij}}$ is the expected number of links in community $s$, given the null model $p_{ij}$, 
and $\gamma$ is a tunable parameter. Minimizing $\h$ defines the community structure.  
When $\gamma=1$,  Eq. (\ref{eq:rb1}) becomes Eq. (\ref{eq:newman3}) apart from a constant factor. 
Hence the RB method contains the modularity 
optimization as a special case, and can be viewed in a more general framework. 
Changing $\gamma$ allows to explore the community structure at different resolutions, but
communities with large differences in size cannot be simultaneously detected using a single value of $\gamma$~\cite{RefWorks:105}.

Recently Arenas, Fern\'andez and G\'omez (AFG) proposed a method~\cite{RefWorks:110} for augmenting modularity 
optimization with a parameter, which similarly to $\gamma$ above allows tuning the resolution of the method. 
This approach considers the network to be weighted. 
The trick introduced by Arenas \emph{et al.}~\cite{RefWorks:110} is to add a self-link of weight $r$ to each 
node, in which case  
the modularity becomes 
\be 
Q_w(r) = \frac 1 {W(r)} \sum_{s=1}^m \left( w_{ss}(r) - [w_{ss}(r)]  \right),
\label{eq:afg5} 
\ee 
where $W(r)$ is total link weight in the network (including self-links),
$w_{ss}(r)$ is total link weight inside community $s$ and $[w_{ss}]$ is its expected value.
Parameter $r$ adjusts the total weight in the network, which in turn 
changes the community detection resolution~\cite{RefWorks:110}. 
Sweeping $r$ and observing which communities are most stable with respect to changes in $r$ should
reveal the community structure.  
 
Eqs.~(\ref{eq:rb1}) and (\ref{eq:afg5}) suggest that RB and AFG methods are somewhat related, although not equal.  
The tuning parameters, $\gamma$ and $r$, behave qualitatively in the same way:  
large parameter values allow finding small communities, and small values yield large communities.  
In fact, in the RB method, the effect of $\gamma$ in Eq.(\ref{eq:rb1}) can be interpreted such that  
the "effective" number of links in the network equals $L/\gamma$,
whereas the parameter $r$ in Eq.~(\ref{eq:afg5}) changes the total weight in the network. 
However, there is a difference: $r$ also increases  
the sum of weights within a community, whereas $\gamma$ has no effect on the number of links within a community.  
In order to illustrate the connection between these methods, we next derive the ``resolution limit'' 
intrinsic for Eq.~(\ref{eq:afg5}) in the AFG method. 
 
Now suppose that a network consists of "physical" communities, which are somehow known to us. We consider two 
of these communities, $s$ and $t$, such that the sum of weights of edges connecting them is $w_{st}$. 
If these "physical" communities are merged by the detection method, the modularity $Q_w(r)$ changes by 
$\Delta Q_w(r)  =  \frac 1 {W(r)} \left( w_{st} - [w_{st}(r)] \right)$.
The optimization of modularity should merge these communities if $\Delta Q_w(r) > 0$, which yields 
\be 
S_s(r)S_t(r)< 2W(r) w_{st},
 \label{eq:afg7} 
\ee 
where $S_s(r)$ is the total node strength in community $s$.
An analogous result for  RB method is $\gamma K_s K_t < 2L l_{st}$, where $K_s$ is total 
node degree in community $s$.
Hence the tuning parameters $\gamma$ and $r$ are not identical, and they affect the optimization 
outcome differently. 
However, if we assume that $S_s=S_t\approx n_s \ave s$, $n_s=n_t$ and $K_s \approx n_s \ave k$
Eq.~(\ref{eq:afg7}) reduces to 
\be 
n_s <  \sqrt{ \frac{ N w_{st}}{\ave s + r}}, 
\ee 
which bears resemblance to the corresponding RB result: $n_s < \sqrt{N l_{st}/(\gamma \ave k)}$.
 
Next, we present some numerical results obtained by sweeping the tuning parameters $\gamma$ and 
$r$ of the RB and AFG methods across a range of values, and optimizing the respective energy 
functions using simulated annealing. Three different test networks are used. 
We  show the behavior of the number of communities detected by the methods 
as a function of the tuning parameter, and look for "stable" regions where this number remains constant~\cite{RefWorks:110}. 
Earlier, community structures detected using several values of $\gamma$ in the RB method have been 
reported in~\cite{RefWorks:50}, but to our knowledge complete sweeps and stability 
analysis have not been reported earlier.  
 
\begin{figure}[t] 
\includegraphics[width=1.0\linewidth]{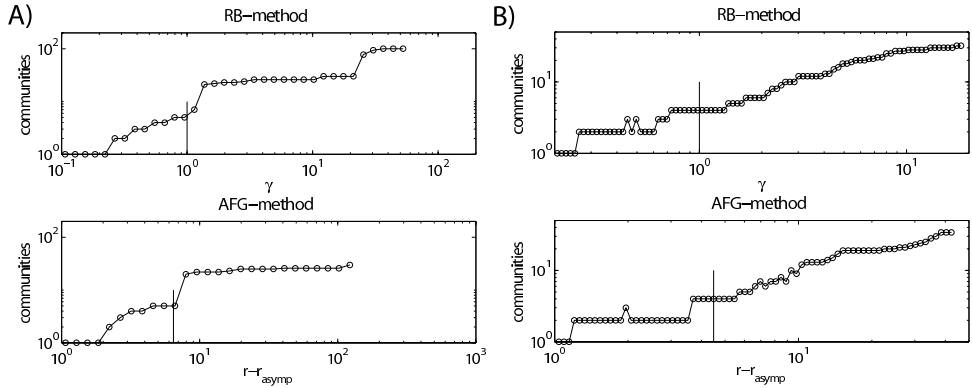} 
\caption{Number of communities as detected with simulated annealing using the RB (upper) and AFG (lower) methods.  
A: hierarchical scale-free network~\cite{RefWorks:118} of 125 nodes, B: Zachary's karate club. 
The vertical line denotes the traditional modularity optimization case. } 
\label{fig:esim1} 
\end{figure} 
 
Our first test network is a synthetic, hierarchical scale-free network of $N=125$ nodes~\cite{RefWorks:118}. 
This unweighted network can be viewed to consist of 5 communities 
of 25 nodes each, which can be further divided into five-node cliques (for a visualization of this network, 
see ~\cite{RefWorks:118} or ~\cite{RefWorks:110}).  
Figure \ref{fig:esim1}(A) shows the number of communities detected using the RB and AFG methods.  
Both  methods are able to reveal the large communities at small values of sweeping parameter, 
although the AFG method seems to perform slightly better. 
One should note that this might be a feature 
of the numerical optimization, and not the method itself. 
We remind the reader that the "traditional" modularity optimization corresponds to $\gamma=1$ and $r=0$. 
These points are shown in the figures as vertical lines. Our results for the AFG method are consistent 
with those reported in ~\cite{RefWorks:110}. 
 
Our second test network is a small, unweighted network representing Zachary's karate club~\cite{RefWorks:116}, 
which has often been used as a "testbed" for community detection. Modularity optimization is known to 
yield four communities, whereas this club was observed to split into two communities. In ~\cite{RefWorks:110},   
the authors demonstrated that AFG method is able to find exactly those communities (by using the weighted 
version of this network). Results for the unweighted network in Fig.~\ref{fig:esim1}(B) show that both methods 
give similar results and are able to detect the two communities. A closer inspection shows that the communities 
correspond to the split which eventually happened (except for one individual classified differently by the RB method). 
 
\begin{figure}
 \begin{center}
   \subfigure[] {\includegraphics[width=0.45\linewidth]{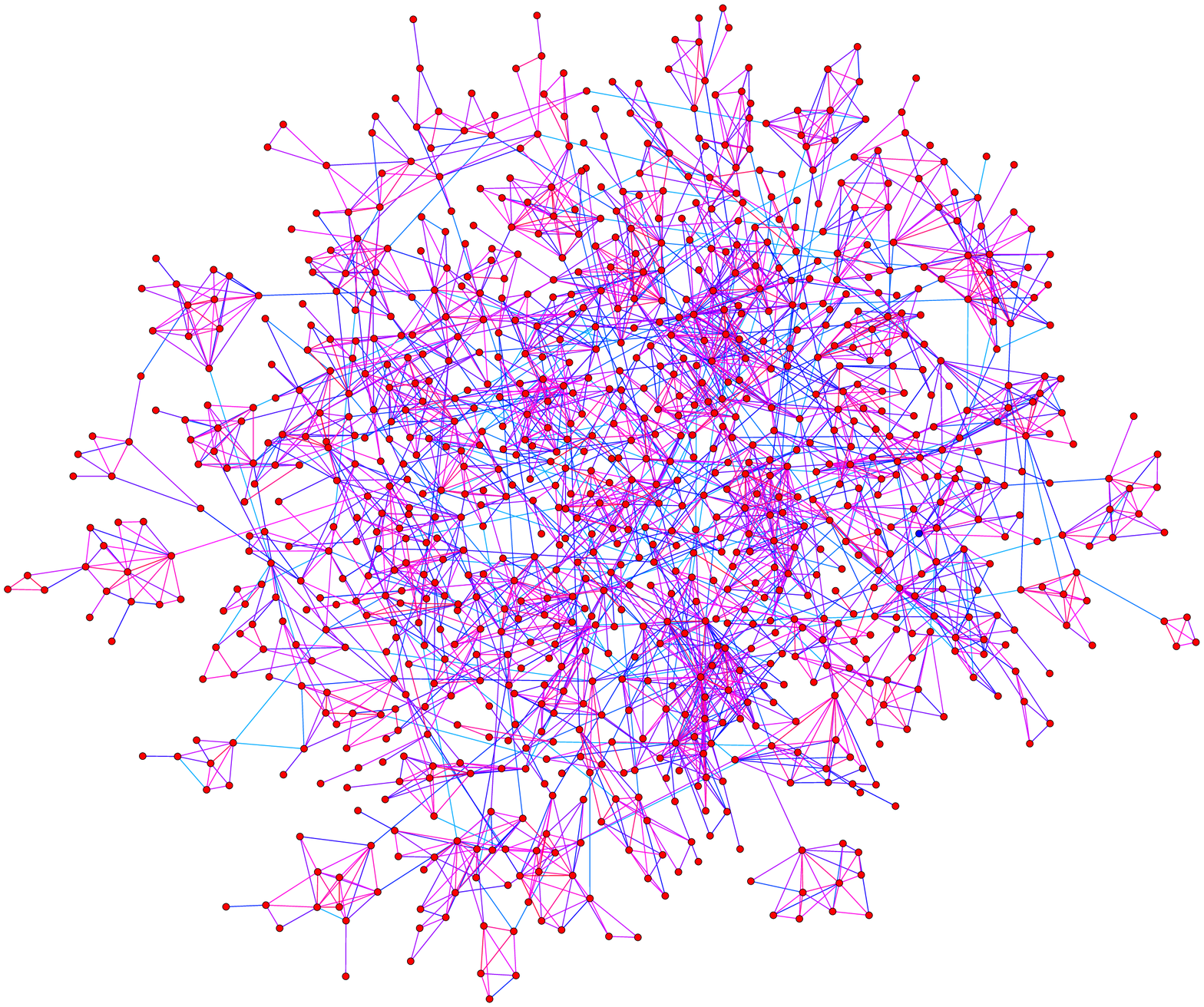} }
   \subfigure[] { \includegraphics[width=0.45\linewidth]{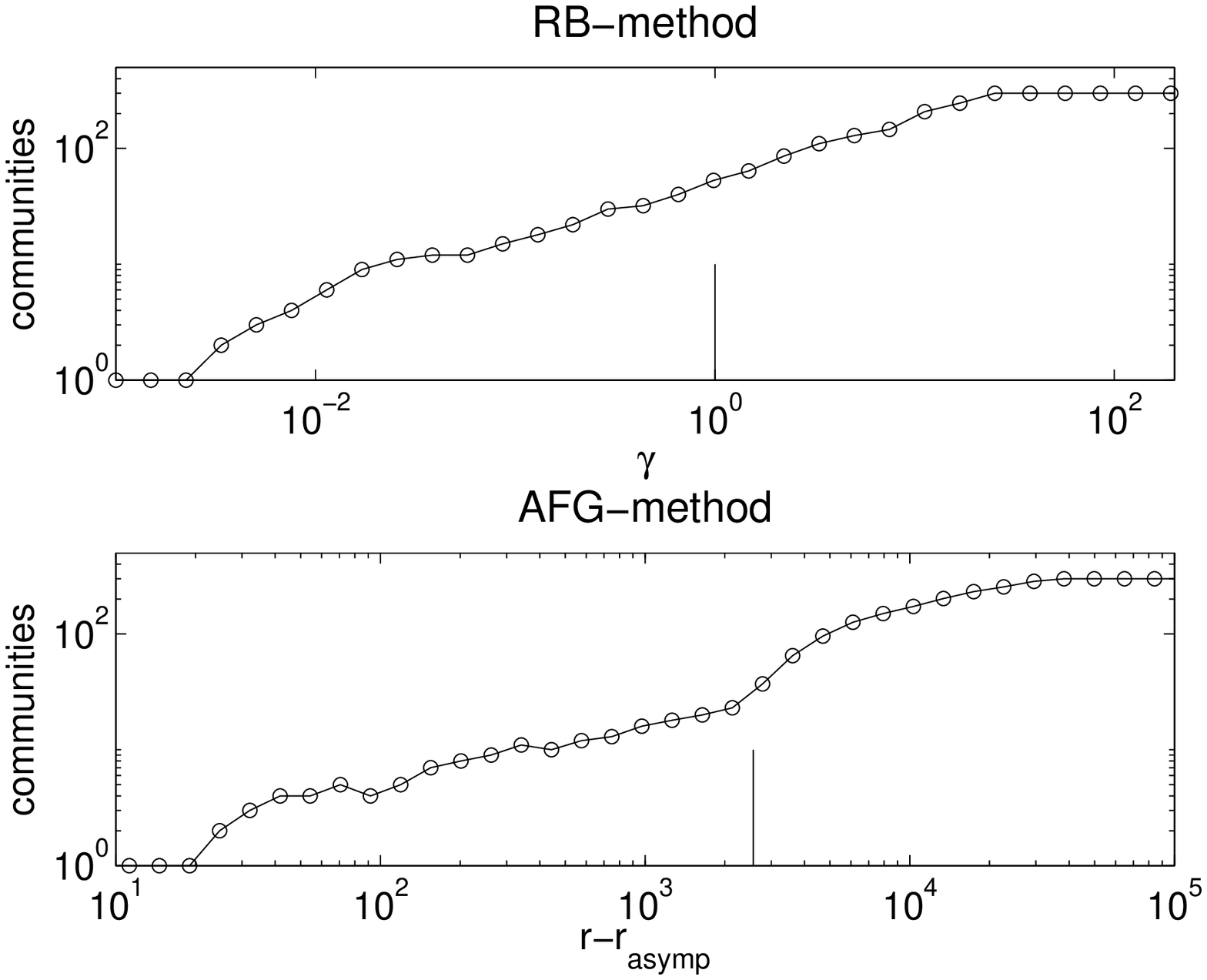} }
   \caption{  (Color online) A weighted test network having 986 nodes. Link colors vary from blue (weak) to red (strong), 
 Number of communities for the network as a function 
of the tuning parameters. Note that we have limited the number of communities to 300. }
    \label{fig:esim4}
  \end{center}
\end{figure}

Our third test network is weighted, being larger than the previous examples (986 nodes), and has a more complex 
community structure, Fig.~\ref{fig:esim4}(a). The average degree of this network is $\ave k=6$ and it  
has been generated with a model designed to resemble real, weighted social networks. 
Visually, the communities are less apparent than in the previous test networks, 
although it can be seen that there are dense groups of nodes with strong internal links, connected 
by weaker links.  Applying the clique percolation method~\cite{RefWorks:52,RefWorks:113,RefWorks:114} 
to this network using clique size 4 yields communities whose sizes vary from 4 nodes (20 communities) 
to 43 nodes (1 community).  Because the network is weighted, we have used the a weighted Hamiltonian
instead of (\ref{eq:rb1}) for the RB method \cite{RefWorks:105}. 
Results in Fig. \ref{fig:esim4}(b) show that no clear "stable" regions of the tuning 
parameters with a constant number of communities are apparent. 
One possible explanation is that this is due to quite 
non-uniform distribution of community sizes, which may result in large communities continuously being 
split into smaller ones as the tuning parameters are increased. A similar situation could occur  
for many large real-world networks.  However, by using  small values of $\gamma$ and $r$ it might 
be possible to study the large-scale community structure, such that the network is split into  
a small number of large communities. 
 
We have discussed the limited resolution of community detection methods where a global energy-like  
 quantity is optimized, by focusing especially on two methods (RB and AFG)  
where the resolution can be adjusted using a tuning parameter. Although the tuning parameters  
of these two methods give rise to qualitatively similar changes in resolution, analytic derivations 
show that their effect on the resolution limit is somewhat different. These two methods have also 
been numerically tested by using simulated annealing, with the result that in small test networks, 
stable regions of tuning parameter values, where the number of communities is constant, can easily 
be found. These can be viewed to reflect "optimal" communities. However, on a large, 
weighted test network, where the clique percolation method indicates a broader distribution of 
community sizes, no such regions are apparent.

{\bf Acknowledgments:} This work was partially supported by the Academy of Finland  
(Center of Excellence program 2006-2011). JS acknowledges support by the European 
Commission NEST Pathfinder initiative on Complexity through project EDEN 
(Contract 043251). JK is partly supported by OTKA K60456.


\end{document}